\documentclass{article}[12pt]%

\usepackage{url}
\usepackage{tikz}
\usepackage{natbib}

\usepackage{a4wide}
\usepackage[margin=.5cm]{caption}

\usepackage{amsmath}
\usepackage{bm}
\usepackage{graphicx}
\usepackage{changes}
\setdeletedmarkup{{\textcolor{gray}{\sout{#1}}}}
\setaddedmarkup{{\textcolor{red!80!black}{#1}}}
\definechangesauthor{MAC}
\definechangesauthor{MDW}
\definechangesauthor{JGS}

\usepackage{pifont}

\newcounter{editnumber}

\def\id{\mathbf{I}}
\newcommand\etal{\textit{et al.}}

\newcommand\compactlist{\setlength{\itemsep}{0pt}\setlength{\parskip}{0pt}}

\begin{document}

\title{Exploring the consequences of lack of closure in codon models}

\author{Michael D.\ Woodhams$^{1\ast}$, Jeremy G.\ Sumner$^{1}$, David A.\ Liberles$^{2}$,\\
Michael A.\ Charleston$^{1,3}$,  Barbara R.\ Holland$^{1}$}


\maketitle

{$^{1}$University of Tasmania, TAS, Australia, $^\ast$email: michael.woodhams@utas.edu.au; \\\indent$^{2}$Temple University, Philadelphia, PA, USA; $^{3}$University of Sydney, NSW, Australia }

\abstract{\noindent Models of codon evolution are commonly used to identify positive selection. 
Positive selection is typically a heterogeneous process, i.e., it acts on some branches of the evolutionary tree and not others. 
Previous work on DNA models (Sumner et al. 2012, {\em Is the General-Time-Reversible model bad for molecular phylogenetics?}) showed that when evolution occurs under a heterogeneous process it is important to consider the property of model closure, because non-closed models can give biased estimates of evolutionary processes. 
The existing codon models that account for the genetic code are not closed; to establish this it is enough to show that they are not linear (meaning that the sum of two codon rate matrices in the model is not a matrix in the model). 
This raises the concern that a single codon model fit to a heterogeneous process might mis-estimate both the effect of selection and branch lengths. \\
Codon models are typically constructed by choosing an underlying DNA model (e.g., HKY) that acts identically and independently at each codon position, and then applying the genetic code via the parameter $\omega$ to modify the rate of transitions between codons that code for different amino acids. 
Here we use simulation to investigate the accuracy of estimation of both the selection parameter $\omega$ and branch lengths in cases where the underlying DNA process is heterogeneous but $\omega$ is constant. 
We find that both $\omega$ and branch lengths can be mis-estimated in these scenarios. 
Errors in $\omega$ were usually less than 2\% but could be as high as 17\%.
We also assessed if choosing different underlying DNA models had any affect on accuracy, in particular we assessed if using closed DNA models gave any advantage. 
However, a DNA model being closed does not imply that the codon model constructed from it is closed, and in general we found that using closed DNA models did not decrease errors in the estimation of $\omega$.
}

\medskip


\section{Introduction}


One of the major goals of evolutionary genomics is the identification of genomic changes that underpin phenotypic differences between species \cite{anisimova2012detecting}. 
To identify such changes it is important to identify regions or even individual sites in molecular sequences that have been under differing selective pressure through evolutionary history. While some of these phenotype-inducing changes occur in regulatory regions that are not responsible for encoding protein sequences, a large subset will change the amino acid sequence of a protein \citep{barrett2011molecular}. 
At the interspecific level, several classes of codon models have been described that enable detection of changes in selective pressure in protein encoding genes across a phylogeny.  

Markov models of codon substitution were introduced by \citet{goldman1994codon} and \citet{muse1994likelihood} and have subsequently been extended so that the ratio of nonsynonymous to synonymous nucleotide substitution rates can be estimated over branches of a phylogenetic tree \citep{yang1998likelihood} or on specific sites undergoing positive selection on particular branches \citep{Zhang2005}. 

In this discussion, we use the term ``model'' to mean a set of substitution matrices, and their corresponding rate matrices, which are parametrized in a particular way, such as the ``Jukes-Cantor'' (JC) DNA model \citep{jukes1969evolution} (which only has one parameter and therefore exactly one rate matrix up to scaling) or the ``general time reversible'' (GTR) DNA model \citep{tavare1986some} (which has nine free parameters).
Further, when we say a matrix is ``in a model'', we mean that its entries, be they substitution probabilities or rates,  are obtainable by some parametrization of that model.

When codon models are used to detect positive selection, they allow selection to differ between branches.
For instance, the parameter $\omega$ that controls the ratio of non-synonymous to synonymous changes might be free to vary in different branches \citep{yang1998likelihood}.
Recent work by Sumner \etal\ (2012) has shown that modelling a heterogeneous process as homogeneous can be inaccurate if the underlying substitution models do not have the mathematical property of (multiplicative) {\em closure}.
Mathematically, closure is the idea that if one multiplies two substitution matrices from some model, then their product is also in the same model.
For instance, it is known that, in general, two GTR substitution matrices, when multiplied together, need not yield another GTR matrix; on the other hand, the product of two F81 substitution matrices \cite{felsenstein1981} is always an F81 matrix.
Both results are demonstrated in \cite{sumner2012lie}.
From this it is clear that some of the standard phylogenetic models currently in use are closed and some are not.
To rectify this situation a complete hierarchy of closed DNA models, coined  Lie-Markov models, has been developed; in particular allowing for distinguished nucleotide pairings \citep{fernandez2015lie, woodhams2015new}.

In this paper we focus on generalisations of two different, but closely related, types of codon model: those introduced by \citet{muse1994likelihood} (MG-type models), and those introduced by \citet{yang1998synonymous} and \citet{nielsen1998likelihood} derived from \citet{goldman1994codon} (GY-type models).
In these codon models DNA substitution rates are scaled by either 0, 1, or $\omega$, depending on whether, following the genetic code, the codon substitution is prohibited, or induces an amino acid substitution that is synonymous or non-synonymous, respectively.
To distinguish from DNA substitution rates and probabilities, we will refer to the parameters that arise from the genetic code as ``augmentation'' parameters, comprising both selection parameter $\omega$ and the existence of stop codons.

As discussed in \citet{lindsay2008pitfalls}, following the GTR paradigm of model construction, both MG- and GY-type models can be decomposed into a symmetric relative rate component and a base-frequency component. 
The difference is that, for example, if codon AAA goes to AAG, in the GY-type models the overall rate depends on the frequency of the codon AAG whereas in the MG-type models it only depends on the frequency of the nucleotide G. 
For reasons detailed below, the MG-type models have a more sensible mathematical structure, in the sense that if the underlying DNA model has the closure property then this at least carries over to the resulting MG-type \emph{triplet} model (see later), which can be thought of as the MG-type \emph{codon} model without the augmentation parameters. Further, MG-type models are better justified biologically from the perspective that the same DNA-level mutational process is likely to act at each of the three codon positions, with only selection differing. This is born out in their providing less biased estimates of selective parameters \cite{Spielman2015}. 
This property is not satisfied by the GY-type models because the multiplication of the relative rates with the codon frequencies necessarily introduces non-linear constraints.


For DNA models, \citet{sumner2012general} found that lack of closure can cause serious mis-estimation of model parameters.
In the case of codon models, there are two important and independent reasons why they do not have the closure property.
The first is that codon models are built up from DNA models acting independently at each codon position, and these underlying models need not themselves be closed.
The second is due to the augmentation parameters 0, 1, $\omega$, used to reflect the structure of the genetic code.
In the section that follows we will tease apart these two effects and formally establish the lack of closure for codon models.

The lack of closure raises a somewhat alarming prospect, that the resulting artefacts could cause  mis-estimation of the parameters used to understand selection.
To make this concrete, consider a scenario where for a period of time one substitution matrix $M_1$ governs codon evolution, which is then followed by a period governed by a distinct substitution matrix $M_2$.
Suppose also that selective pressure has not changed, i.e., $\omega$ is the same in both matrices, but there has been a change in the underlying DNA substitution rates.
We need not assume any change in the underlying DNA model, only in its substitution rate parameter values.
Even with these simple assumptions, the combined process, given by the matrix product $M_1M_2$, is not, in general, obtainable using the same model.
However, there will be some alternative codon substitution matrix $M'$ in the model that \emph{best fits} the observed data, and the $\omega$ parameter in the ``compromise'' $M'$ may not be the same as the $\omega$ in models $M_1$ and $M_2$.

In this paper, we seek to evaluate whether the lack of closure of codon models, be it caused by the underlying DNA model and/or by the augmentation parameters, has a significant detrimental effect on estimation of model parameters. 
We simulate cases where the DNA rate parameters differ over two lineages while $\omega$ remains constant.
We measure the resulting error in estimation of both branch lengths and $\omega$, and explore whether the use of closed DNA models reduces these errors.

This setup explores the errors that can arise when fitting a homogeneous model to what is truly a heterogeneous process.
For clarity, since we allow for DNA rate parameters to differ \emph{across} lineages and then attempt a homogeneous fit, the model specification is \emph{not} necessarily cured by applying a multiplicatively closed model.
Our results are consistent with this: whether applying time reversible or multiplicative closed underlying DNA models in a codon model, we obtain errors in $\omega$.  
We do however frame our discussion in the context of multiplicative closure, since the scale of the errors is due to the non-linearity present in the parametrization of these models and it is this property that is brought to light by consideration of multiplicative closure.
 
Additionally, the attraction of multiplicatively closed models is that they can be consistently used when the process is heterogeneous along a single lineage, and thus can be used to consistently fit a heterogeneous model to a true heterogeneous processes on a tree, even under the presence of subsampling of taxa.
This consistency is not shared by most of the family of time-reversible DNA models or any of the presently available codon models.

This motivates the need for development of closed models, similar to the newly introduced hierarchy of Lie-Markov models for DNA \citep{fernandez2015lie, woodhams2015new}, that apply at a \emph{codon} level. 
We conclude the paper by giving some first thoughts on what such models might look like.

\section{Closure properties for codon models}

We are interested in drawing on algebraic properties of substitution models to better understand closure properties, or lack thereof, of both MG- and GY-type models.
To properly make these mathematical connections we present a generalised formulation in what follows.
In particular, we will follow the framework of embedding the models within the general Markov model of codon substitutions, making connections to the GTR paradigm when needed for further understanding.

Our approach will be to first construct what we will call a \emph{triplet} model by assuming independence of the substitutions at the three nucleotide positions, as is typical in practice.
We use ``triplet model'' to distinguish it from the more complex codon models, which include the augmentation parameters.
We then modify the triplet model to account for synonymous and nonsynonymous amino acid substitutions and stop codons, as dictated by the genetic code, yielding a \emph{codon} model.
We will see how the first part of the construction can be understood in algebraic means that are perfectly compatible with the goal of producing a multiplicatively closed model, whereas the second part, involving the augmentation parameters, is incompatible with that goal.
In this way, we are led to a deeper understanding of the obstructions involved in producing a multiplicatively closed codon model that respects the genetic code.

Suppose $M$, $N$, and $P$ are $4\times 4$ DNA substitution matrices whose entries give the probabilities of DNA substitution at nucleotide positions 1, 2, and 3 respectively.
Assuming substitutions at different positions are independent, it follows that the probability of transition from triplet $xyz$ to triplet $x'y'z'$ is given by the product
\[
\text{prob}(xyz \rightarrow x'y'z') = M_{xx'}N_{yy'}P_{zz'}.
\]

This array of numbers can be organised using a matrix ``Kronecker'' product $\otimes$, so $\text{prob}(xyz \rightarrow x'y'z')$ is equal to the corresponding entry of the $64\times 64$ matrix $M\otimes N\otimes P$.
The Kronecker product of two matrices $D$, $E$ is obtained by replacing each entry in $D$ by that entry multiplied by the matrix $E$.
To illustrate, the Kronecker product of two $2\times 2$ matrices 
\[
D =\left[\begin{array}{cc} 1 & 0 \\ 1 & -2 \end{array}\right],
\quad 
E = \left[\begin{array}{cc} 6 & -3 \\ 0 & -1\end{array}\right] 
\]
is given by
\[\arraycolsep=5pt
D\otimes E =\left[\begin{array}{c@{~}c} E & 0 \\ E & -2E \end{array}\right]= \left[\begin{array}{c@{~}c@{~}c@{~~}c} 6 & -3 & 0 & 0 \\ 0 & -1 & 0 & 0 \\ 6 & -3 & -12 & 6\\ 0 & -1 & 0 & 2\end{array}\right].
\]

Performing this Kronecker product on the three $4\times 4$ DNA substitution matrices $M$, $N$ and $P$ yields the $4^3\times 4^3 = 64\times 64$ triplet substitution matrix $M\otimes N\otimes P$, where the ordering of triplets is implicitly determined by the properties of the Kronecker product. 
Turning to a continuous time formulation, each substitution matrix is the matrix exponential of some (but, to remain general, possibly different) rate matrix, i.e., $M(t)=e^{At}$, $N(t)=e^{Bt}$, and $P(t)=e^{Ct}$ where $t$ is time.
We can recover the DNA rate matrix, $A$, via the derivative of $M(t)$ evaluated at $t\!=\!0$, giving $A=\left.\frac{d}{dt}M(t)\right|_{t=0}$.
Correspondingly, we can express the $64\times 64$ rate matrix as
\begin{align}
\label{eq:triplet}
R_{\text{triplet}} & =\left.\frac{d}{dt} (M(t)\otimes N(t)\otimes P(t)) \right|_{t=0} \notag \\ & =A\otimes \id \otimes \id +\id \otimes B \otimes \id +\id \otimes \id \otimes C,
\end{align}
where $\id$ is the $4 \times 4$ identity matrix.
(This result follows as an implementation of the usual rule for the derivative of a product.)
This formulation also makes intuitive sense, since the independence assumption implies that the instantaneous rate of substitution between any two triplets with variation at more than one position should be zero, and this can be confirmed by checking the matrix entries of each summand above.

Indeed, one finds that
\[
\text{rate}(xyz \rightarrow x'y'z')=
\left\{
\begin{array}{cl}
A_{xx'}, & \text{ if }y=y',z=z';\\
B_{yy'}, & \text{ if }x=x',z=z';\\
C_{zz'}, & \text{ if }x=x',y=y';\\
0, & \text{ otherwise},
\end{array}
\right.
\]
which is consistent with the structure of MG-type codon models.

Following the basic properties of Kronecker products, a simple argument then shows that such a triplet model is closed if, and only if, each rate matrix $A,B,C$ belongs to a (possibly different) closed DNA substitution model (see Supplementary Material).

Given this generalized version of triplet models, we now turn to accounting for the genetic code using the augmentation parameters.
We refer to these models as \emph{codon} models.
We like to describe the introduction of the augmentation parameters, in particular, the dN/dS ratio $\omega$, into the codon rate matrix as an ``overlay matrix''.
We create the overlay matrix by encoding the genetic code into a $64\times 64$ matrix $G$ where each entry is either a $0, 1,$ or $\omega$, depending on whether the corresponding substitution is prohibited (stop codons), synonymous, or nonsynonymous, respectively.
A substitution is prohibited if it is to or from a stop codon.
We note that it is not necessary to explicitly prohibit multiple simultaneous DNA substitutions in the same codon, as these are automatically prohibited by the underling triplet model.
This is a direct reflection of the Kronecker product construction of $R_\text{triplet}$ above.

Our generalized model is then expressed as a simple entry-wise multiplication of the triplet rate matrix $R_{\text{triplet}}$ with the matrix $G$, followed by adjustment of diagonal entries to ensure zero row-sums. 
When needed, we write this two-step process using the notation
\begin{equation}
\label{eq:codonG}
R_\text{codon}=R_\text{triplet}\ast G,
\end{equation}
where the off-diagonal $i,j$-th entry in  $R_\text{codon}$ is given by the product of the $i,j$-th entries of $R_\text{triplet}$ and $G$, and the diagonal entries are determined by demanding zero row-sums.
We note that this codon model is equivalent to the General Nucleotide Codon model of \citet{kaehler2017standard}.

We refer to any codon model constructed in this way as being of \emph{MG-type}, since, in the special case where the underlying DNA rate matrices are the same and are selected from the F81 model, we obtain precisely the Muse-Gaut codon model \citep{muse1994likelihood}.
This gives a convenient mathematical description of MG-type models but, unlike the Kronecker product operation $\otimes$, the introduction of the augmentation parameters via $G$ and the operation $\ast$ described above does \emph{not} preserve the closure property.
As we will presently establish, the introduction of the augmentation parameters causes the model to become non-linear, which by itself, destroys any chance of multiplicative closure.

Expressed in terms of constraints on rate matrices, for a model to be multiplicatively closed, the set of rate matrices obtainable from the model must form a \emph{linear space} (along with an additional condition not needed here involving `Lie brackets', or `commutators' --- see \citep{sumner2012lie,sumner2017multiplicatively} and also the Supplementary Material for more details).
This linearity condition means that the sum of any two rate matrices from the model is another rate matrix in the model.
Na\"ively (albeit mostly accurately) this property can be detected by inspecting the parametrization of the rates of substitutions: almost always, if the substitution rates are expressed using products of parameters, the model cannot be linear.
For example, the rates for the GTR and HKY models involve products of parameters for the equilibrium distribution and what are sometimes referred to as `relative' rates.
These DNA models are not linear, and hence are not multiplicatively closed.
On the other hand, the rates of substitution in the Kimura 2ST (K2ST) and the F81 models are expressed without products of parameters, and indeed these models are linear.

To illustrate the issue with any codon model constructed as in (\ref{eq:codonG}), suppose the underlying DNA model is identical at each position (that is, $A=B=C$), with, for example, the F81 model selected as the underlying DNA model (recall this is a closed DNA model).
In this DNA model the rate matrix can be expressed as 
\[\arraycolsep=5pt
Q_{\text{F81}} = \left[\begin{array}{c@{~}c@{~}c@{~}c} 
- & \alpha_2 & \alpha_3 & \alpha_4 \\
\alpha_1 & - & \alpha_3 & \alpha_4 \\
\alpha_1 & \alpha_2 & - & \alpha_4 \\
\alpha_1 & \alpha_2 & \alpha_3 & - \\
\end{array} \right],
\]
where `$-$' is a value fixed such that the row sum is 0, together with the triplet rate matrix given by
\[
R_{\text{triplet}}=Q_{\text{F81}}\otimes I\otimes I+I\otimes Q_{\text{F81}}\otimes I+I\otimes I\otimes Q_{\text{F81}}.
\]
The parameters $\alpha_i$ are not normalised in any particular way; however the DNA equilibrium base frequencies $\pi_i$ can be calculated as
\[
\pi_i = \frac{\alpha_i}{\alpha_1 + \alpha_2 + \alpha_3 + \alpha_4}.
\]

Now notice that the distinct entries of the resulting codon rate matrix $R_\text{codon}=R_\text{triplet}\ast G$ in this model are all either 0 (which we can, and will, ignore in the following argument), $\alpha_i$, or $\alpha_j\omega$, depending on whether the substitution is prohibited, synonymous or nonsynonymous, respectively.
If we add two such matrices together then the non-zero entries have the form $\alpha_i+\alpha_i'$ and $\alpha_j\omega+\alpha_j'\omega'$.
For the matrix sum to be an element of our codon model, these new entries must be expressible in the form
\begin{align}
\label{eq:sum}
\widehat{\alpha}_i &= \alpha_i+\alpha_i', \nonumber\\
\widehat{\alpha}_j\widehat{\omega} & = \alpha_j\omega+\alpha_j'\omega'
\end{align}
for some choices of $\widehat{\alpha}$ and $\widehat{\omega}$.

Consistent with all matrices in this codon model, the resulting matrix sum should contain multiple non-linear relationships between rates; in particular, the obvious equality
\begin{equation}
\frac{\widehat{\alpha}_1\widehat{\omega}}{\widehat{\alpha}_1}= \frac{\widehat{\alpha}_2\widehat{\omega}}{\widehat{\alpha}_2}.
\end{equation}
\noindent then produces the following constraint on parameters:
\[
\frac{\alpha_1\omega+\alpha_1'\omega'}{\alpha_1+\alpha_1'} = 
\frac{\alpha_2\omega+\alpha_2'\omega'}{\alpha_2+\alpha_2'},
\]
which is clearly violated for general choices.
Thus, we conclude that this codon model is not closed.

Provided the underlying DNA models themselves have free parameters (beyond an overall scaling), a similar argument establishes that any codon model constructed as in (\ref{eq:codonG}) is not closed. 
Even in the special case where the preceding caveat is false (such as the Jukes-Cantor DNA model) and the resulting codon model is linear, it nonetheless follows that the structure of the genetic code and the augmentation parameters 0, 1, $\omega$ themselves cause the model not to be closed.

We close this section with a discussion of how the GY-type models fall  within the general perspective given thus far.
The major difference with the MG-type models is the treatment, within the GTR framework, of equilibrium frequencies and `relative' rates.
For GY-type models, it is the matrix of relative rates that can be expressed using the Kronecker product operation.
For instance, if we take $S$ as a symmetric $4\times 4$ DNA rate matrix, we can construct a symmetric $64\times 64$ matrix of relative rates by taking:
\[
S_{\text{triplet}}=S\otimes \id \otimes \id +\id \otimes S \otimes \id +\id \otimes \id \otimes S.
\]
(As mentioned above, this construction can be generalized by using a different symmetric DNA rate matrix at each codon position whilst maintaining the codon site independence assumption.)
The GY-type models are then obtained by taking a $64\times 64$ diagonal matrix $D$ of codon equilibrium frequencies (chosen in various ways) and, as is usual in the GTR framework, taking the product {$S_{\text{triplet}}D$ and then introducing the augmentation parameters by defining:
\[
R_{\text{GY}}=S_{\text{triplet}}D\ast G.
\]

In the particular case that each symmetric DNA matrix $S$ is taken from the K2ST model, we obtain the original model used by Goldman and Yang [\citeyear{goldman1994codon}].
As discussed in the introduction, we prefer, and will focus on, the MG-type construction.

\section{Methods}

\subsection{General approach of simulation study}

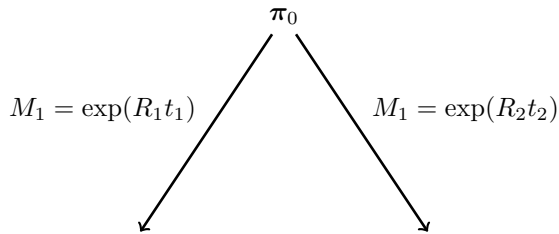
\begin{figure}
\begin{center}
\begin{tikzpicture}[line width=1pt]
\draw(2,3) node(root) {$\bm{\pi}_{0}$} (0,0) node(left) {} (4,0) node(right) {};
\path[->](root) edge [above left] node {$M_{1} = \exp(R_{1}t_{1})$} (left);
\path[->](root) edge [above right] node {$M_{1} = \exp(R_{2}t_{2})$} (right);
\end{tikzpicture}
\end{center}

\caption{
Initial codon frequencies, $\bm{\pi}_0$, are either: (1) the average of the equilibrium codon frequencies given by $M_1$ and $M_2$; or (2) the equilibrium base frequency distribution of a matrix $M$, where $M$ is chosen randomly in the same way as  $M_1$ and $M_2$. 
We then evolve by Markov matrices $M_{1}$ and $M_{2}$ 
\label{fig:GY}}
\end{figure}
As outlined in the introduction, there is potential for the lack of closure of codon models to cause over- or under-estimation of model parameters. 
We wanted to know if fitting a homogeneous codon model to a heterogeneous process could lead to mis-estimation of either $\omega$ or branch lengths in simple scenarios with two taxa and a single change of process. 
We explored what magnitude of errors were generated under a range of biologically reasonable evolutionary scenarios.

We were also interested in the follow-up question of whether or not different choices of underlying DNA model had a significant effect on the magnitude of errors. 
In particular we wanted to test if there was any advantage to constructing codon models from closed (i.e., Lie-Markov) DNA models.

In order to address these questions, we applied the following simulation procedure to MG-type models 
 (see Figure~\ref{fig:GY} for illustration):
\begin{enumerate}\compactlist
	\item Fix a DNA model of interest, e.g., the HKY model. 
	(For a complete list of models tested see Table \ref{tab:model_families}.)
	\item Randomly choose parameters for two DNA rate matrices from this model.
	\item Choose branch lengths $t_1$ and $t_2$ and a fixed value of $\omega$.
	\item Generate the corresponding codon rate matrices $R_1$ and $R_2$ and measure how different they are.
	\item Either set the initial codon frequency distribution $\bm{\pi}_0$ to be the average of the equilibrium frequencies for $R_1$ and $R_2$, or randomly select the initial codon frequencies.
	\item Evolve the codon frequencies on the two branches (one for each of $R_1$, $R_2$) to generate a joint probability matrix $J$.
	\item Treat $J$ as representing the initial and final states of a single MG-type model, and find the best fitting choice of DNA rate matrix $\widehat{Q}$, selection parameter $\widehat{\omega}$ and branch length $\widehat{t}$ .  
	\item Compare the estimated selection parameter $\widehat{\omega}$ and estimated branch length $\widehat{t}$ to the true values $\omega$ and $t_1 + t_2$. 
\end{enumerate}

\subsection{Controlling for the effect of differences in models}

As discussed in the background section, codon models are constructed by first assuming an underlying DNA model. 
When generating our heterogeneous process at the codon level, we begin by selecting two random instances from the same DNA model.

DNA models differ in their numbers of free parameters, e.g., beyond an overall scaling, the Jukes-Cantor model has no free parameters, whereas the HKY model has four (the transition/transversion ratio and three degrees of freedom in the choice of base frequencies).

The base frequencies in the underlying DNA model can have a range of different constraints, from all being the same (no degrees of freedom) to all being permitted to vary (three degrees of freedom); we refer to the Base frequency Degrees of Freedom as the BDF. 
Interestingly, while DNA models that are in common use have either BDF~=~0 (with $\pi_A=\pi_G=\pi_C=\pi_T$) or BDF~=~3 (with $\pi_A,\pi_C,\pi_G,\pi_T$ unconstrained), the hierarchy of Lie-Markov DNA models presented in Woodhams \etal\ (\citeyear{woodhams2015new}) displays two additional options: BDF~=~1 resulting from the constraints $\pi_G=\pi_A$ and $\pi_C=\pi_T$, and BDF~=~2, from the constraint $\pi_G+\pi_A=\frac{1}{2}=\pi_C+\pi_T$.
Hence we also attempt to account for this potential confounding variable by including a range of models with BDF~=~0, 1, 2 and 3 (see Table~\ref{tab:model_families}).

It is important to comment at this point on what the distance between two matrices is. 
The distance between two codon rate matrices is computed by first scaling them so they have trace of -1, and then taking the square root of the sum of squared differences of the off-diagonals.
The potential difference between numbers of free parameters means that when we choose parameters at random it is expected to get larger distances between the codon rate matrices $R_1$ and $R_2$ for some DNA models than for others.  
In order to sensibly compare the performance of different classes of DNA models, e.g., time-reversible (TR) models vs.\ Lie-Markov (LM) models, it is important to control for such potential confounding variables. 
For this reason we chose DNA models with a range of features: number of parameters, base frequency degrees of freedom (BDF), LM or not, and TR or not (see Table~\ref{tab:model_families}). 
For each simulation we also recorded the distance between the $R_1$ and $R_2$ codon rate matrices and the difference in the equilibrium base frequencies.  
We were then able to fit a linear regression analysis with error (in either selection parameter $\widehat{\omega}$ or branch length $\widehat{t}$) as the response variable, and characteristics of the models and simulations as predictors. 
This framework allowed us to statistically  test whether LM models have significantly smaller errors than TR models.

\begin{table*}
\begin{centering}
\begin{tabular}{ccccc}
\hline 
BDF=0 & BDF=1 & BDF=2 & BDF=3 & \#~parameters \tabularnewline
\hline 
\bf\em 2.2b (K2P) & \bf\em 3.4 &  & \em 5.6b & 1\tabularnewline
\bf\em 3.3a (K3P) & \em 4.5a & \em 5.7a & \em 6.7a & 2\tabularnewline
\bf\em 3.3c (TrNef) & \bf\em 4.4b & \em 5.11a & \em 6.8a & 2\tabularnewline
\em 5.6a & \em 6.6 &  & \em 8.10a & 5\tabularnewline
\em 5.11c &  & 6.8b & \em 8.16 & 5\tabularnewline
\em 9.20b (DS) &  &  & \em 12.12 (GM) & 8\tabularnewline
\hline 
\bf\em K2P (2.2b) & \bf K2P+1 & \bf HKY-1 & \bf HKY & 1\tabularnewline
\bf\em TrNef (3.3c) & \bf TrN-2 & \bf TrN-1 & \bf TrN & 2\tabularnewline
\bf TVMef & \bf TVM-2 & \bf TVM-1 & \bf TVM & 4\tabularnewline
\bf SYM & \bf SYM+1 & \bf GTR-1 & \bf GTR & 5\tabularnewline
\hline 
\end{tabular}
\par\end{centering}

\caption{The DNA models considered in this study. 
Within each row, models on the left are sub-models of models to their right, differing only in equilibrium base frequency degrees of freedom (BDF). 
The mid-line separates models in the Lie-Markov hierarchy (above) from models in the time reversible hierarchy (below). 
`\#~parameters' is the degrees of freedom of the BDF~=~0 model.
Models in {\em italic} are multiplicatively closed (Lie-Markov), and models in {\bf bold} are time reversible (some models are both).
\label{tab:model_families}}
\end{table*}

\subsection{Constructing the heterogeneous process}
 
For all instances, we first randomly generated DNA equilibrium base frequencies $\bm{\pi}=[\pi_i]$, then generated a random DNA rate matrix $Q$ in the model such that $Q$ had $\bm{\pi}$ as its equilibrium distribution. 
The procedure for generating $\bm{\pi}$ depends on the BDF of the DNA model, and is explained in the following.

In choosing random DNA rate matrices we wanted to avoid parameters that were  biologically unrealistic.
So that individual base frequencies were more likely to be close to 0.25, we used triangular distributions and ensured that no base could have a frequency less than 0.1.
No random generation was required for BDF~=~0 models as these have $\bm{\pi}=(0.25,0.25,0.25,0.25)$. 
For DNA models with BDF~=~1, $\pi_A$ + $\pi_G$ was chosen from a triangular distribution centred on 0.5 with 0.2 and 0.8 as extremes. 
For DNA models with BDF~=~2, $\pi_A$ and $\pi_C$ were independently chosen from a triangular distribution centred on 0.25 with 0.1 and 0.4 as extremes, with $\pi_G$ = $0.5 - \pi_A$ and $\pi_T$ = $0.5 - \pi_C$.
For BDF~=~3 (unconstrained base frequencies) the procedure for generating the $\pi_i$ is complex and is explained in full in the Supplementary Material. 
However, the basic process was to first generate a random $\bm{\pi}^\prime$ as extreme as possible (i.e., containing at least one zero) and then form a weighted average of $\bm{\pi}^\prime$ and $(0.25,0.25,0.25,0.25)$, where the weight of $\bm{\pi}^\prime$ was chosen from a triangular distribution such that the minimum possible value in the average $\bm{\pi}$ was 0.1.

The symmetric part of the time-reversible models was generated via the method of \emph{basis matrices}, in imitation of the construction of Lie-Markov models presented in \cite{woodhams2015new}. 
Details are given in the Supplementary Material.
Under this construction, both time-reversible and Lie-Markov models have parametrizations where parameters must be in the range $[-1,1]$, and if any parameter is $+1$ or $-1$, then the rate matrix will have a zero entry. 
In other words, if a parameter is on the boundary of allowed values, $Q$ will be on the boundary of being a valid (stochastic) rate matrix.
If all parameters are zero, $Q$ will be the Jukes-Cantor matrix.
In either case (LM or TR), we draw parameters from a triangular distribution in the range $[-0.8,0.8]$, so parameters tend to be close to zero and rate matrices tend to be close to the Jukes-Cantor matrix. 

Given a DNA model, to construct the corresponding MG-type model we require two additional parameters: $\omega$ and $t$. 
In our simulations, $\omega$ was fixed and chosen from $\{0.2, 0.5, 1, 1.5, 2\}$, and time $t$ was selected uniformly in the range [0.03,0.18]. 
We have found that these parameter values create conditions in which it is feasible to compare fairly the performance of phylogenetic inference using the resulting simulated data.

Following (\ref{eq:codonG}), the codon substitution matrix $M_i$ for the $i$-th branch is given by
\begin{eqnarray}
\label{eq:M}
M_i=\exp\left(R_i t_i \right)  
\end{eqnarray}
where $R_i=\left(Q_i\otimes \id \otimes \id +\id \otimes Q_i \otimes \id +\id \otimes \id \otimes Q_i\right)\ast G$, and $Q_i$ is the DNA rate matrix and $t_i$ the time on the $i$-th branch.

Having randomly selected codon rate matrices $R_1$ and $R_2$, we then explored two different methods for setting the codon equilibrium frequency $\bm{\pi}_0$ at the root. 
In the first set of simulations we set $\bm{\pi}_0$ to be the average of the equilibrium frequencies for $R_1$ and $R_2$; in the second set of simulations we chose $\bm{\pi}_0$ at random as described above.
In either case, we then generated a joint probability matrix $J$ for the overall process by setting the initial codon frequencies to $\bm{\pi}_0$ and evolving down each branch, using the rate matrices $R_1$ and $R_2$ respectively (Figure \ref{fig:GY}).
$J$ is calculated by
\[
J = M_{1}^{T}\cdot \mbox{diag}(\bm{\pi}_{0})\cdot M_{2}\\
\]
where $M_{i}$ is constructed according to (\ref{eq:M}) from a single DNA rate matrix $Q_i$ that applies at each codon position and which has randomly chosen parameters $\theta_i$.
The $i,j$-th entry of $J$ is the probability that at an arbitrary site, the left leaf has codon $i$ and the right leaf has codon $j$.

\subsection{Model-fitting and performance measures}

We now treat the matrix $J$ as our ``data'' and attempt to fit a single homogeneous process to it. 
The fact that the codon models are not closed means that we will not be able to do this exactly, but we can look for the best fitting homogeneous model. 
We optimize $\omega$, a single model $\mathcal{M}$ with parameters $\theta_\mathcal{M}$ and $t$ to best match $J^\ast$ to $J$, where $J^\ast$ is calculated as follows:
\begin{equation*}
\begin{split}
J^\ast(\omega,\theta_\mathcal{M},t) & =  \exp(R(\omega,\theta_\mathcal{M})t)^{T}\\ & \qquad \times\mbox{diag}(\mbox{eqbm}(R(\omega,\theta_\mathcal{M}))) \\ & \qquad \times\exp(R(\omega,\theta_\mathcal{M})t)
\end{split}
\end{equation*}
\noindent where $\text{eqbm}(R(\omega,\theta_\mathcal{M}))$ is the equilibrium distribution of $R(\omega,\theta_\mathcal{M})$.

For each simulation, we calculate $\Delta\pi$, a measure of the difference in codon frequencies at the two leaves. This is the root square difference in codon frequencies, i.e.,
\[
\Delta\pi = ||\bm{\pi}_1 - \bm{\pi}_2||_{2} = \sqrt{(\bm{\pi}_1-\bm{\pi}_2)\cdot(\bm{\pi}_1-\bm{\pi}_2)^T}
\]
where, taking $\bm{e}$ as the 64-long vector of ones, $\bm{\pi}_1=\bm{e}\cdot J$, and $\bm{\pi}_2=\bm{e}\cdot J^T$ (the column and row marginalizations of $J$ respectively).

The optimization is done by maximum likelihood, where the log-likelihood is
\[
\log(L(\omega,\theta_\mathcal{M},t|J))\propto\sum_{i,j}J_{ij}\log(J^\ast(\omega,\theta_\mathcal{M},t)_{ij})
\] and we denote $\widehat{\omega}$, $\widehat{\theta}_\mathcal{M}$, $\widehat{t}$ as the parameters which maximize this log-likelihood.

If the GY model is robust to inhomogeneity, we should find $\widehat{\omega} \approx \omega$ and $2\widehat{t} \approx t_1+t_2$. 

Simulations were completed for each model shown in Table~\ref{tab:model_families}, for values of $\omega \in \{0.2, 0.5, 1, 1.5, 2\}$. 
500 replicates were performed for each parameter set. 
For each simulation we record both a raw error and a relative error. 
Raw error is recorded as $\widehat{\omega} - \omega$ and $\widehat{t} - (t_1 + t_2)$, and relative error is defined as the absolute value of the raw error divided by the true value.

\section{Results}

\subsection{Does lack of closure lead to mis-estimation of $\omega$ or branch lengths?}
 
In this subsection we report results for the MG-style codon model that embeds HKY as a DNA model (the most widely used software, PAML, \citep{yang2007paml} for fitting codon models is based on the HKY DNA model). Here the lack of closure results from a change in model parameters such as equilibrium frequencies of the bases. Biologically, this is justified by several phenomena, including  shifts in environmental temperature in bacteria \citep{groussin2011adaptation}, site-wise shifts in amino acid frequencies \citep{pollock2012amino}, and laterally transferred genes that show shifts in base frequencies between the old genome and the new genome \citep{daubin2003source}.
Our results for this model show that lack of closure does cause $\omega$ to be mis-estimated (Figure \ref{fig:error_MGHKY_nasty}(a), Table \ref{tab:mean_error_omega_nasty}; Supplementary Figure~S1, Supplementary Table~S4). 
Over- or under-estimation seems about equally likely for values of true $\omega$ less than or equal to 1, but $\omega$ is more likely to be overestimated for true $\omega$ values of 1.5 and 2. 
Mis-estimation is usually less than 2\%, but can be larger (the maximum relative error was 11.6\% in simulations with $\bm{\pi}_0$ chosen to be intermediate, and 16.9\% in the simulations where $\bm{\pi}_0$ was chosen randomly).

\begin{table*}
\begin{centering}
\begin{tabular}{cccccc}
\hline 
true $\omega$ & 0.2 & 0.5 & 1 & 1.5 & 2 \tabularnewline
\hline 
Mean raw error	& -0.001065 & -0.002390 & -0.003419 & 0.007429 &  0.014113 \tabularnewline
Mean relative error	& 0.018394 & 0.018873 & 0.017968 & 0.016775 & 0.017175 \tabularnewline
Maximum relative error & 0.126965 & 0.168885 & 0.139984 & 0.134708 & 0.169300 \tabularnewline
\hline 
\end{tabular}
\par\end{centering}
\caption{Mean raw error, mean relative error and maximum relative error in the estimate of $\omega$ for different true values of $\omega$ for simulations where $\bm{\pi}_0$ is chosen at random. 
\label{tab:mean_error_omega_nasty}}
\end{table*}

\begin{figure*}
\begin{center}
\begin{tabular}{cc}
\includegraphics[width=0.49\textwidth]{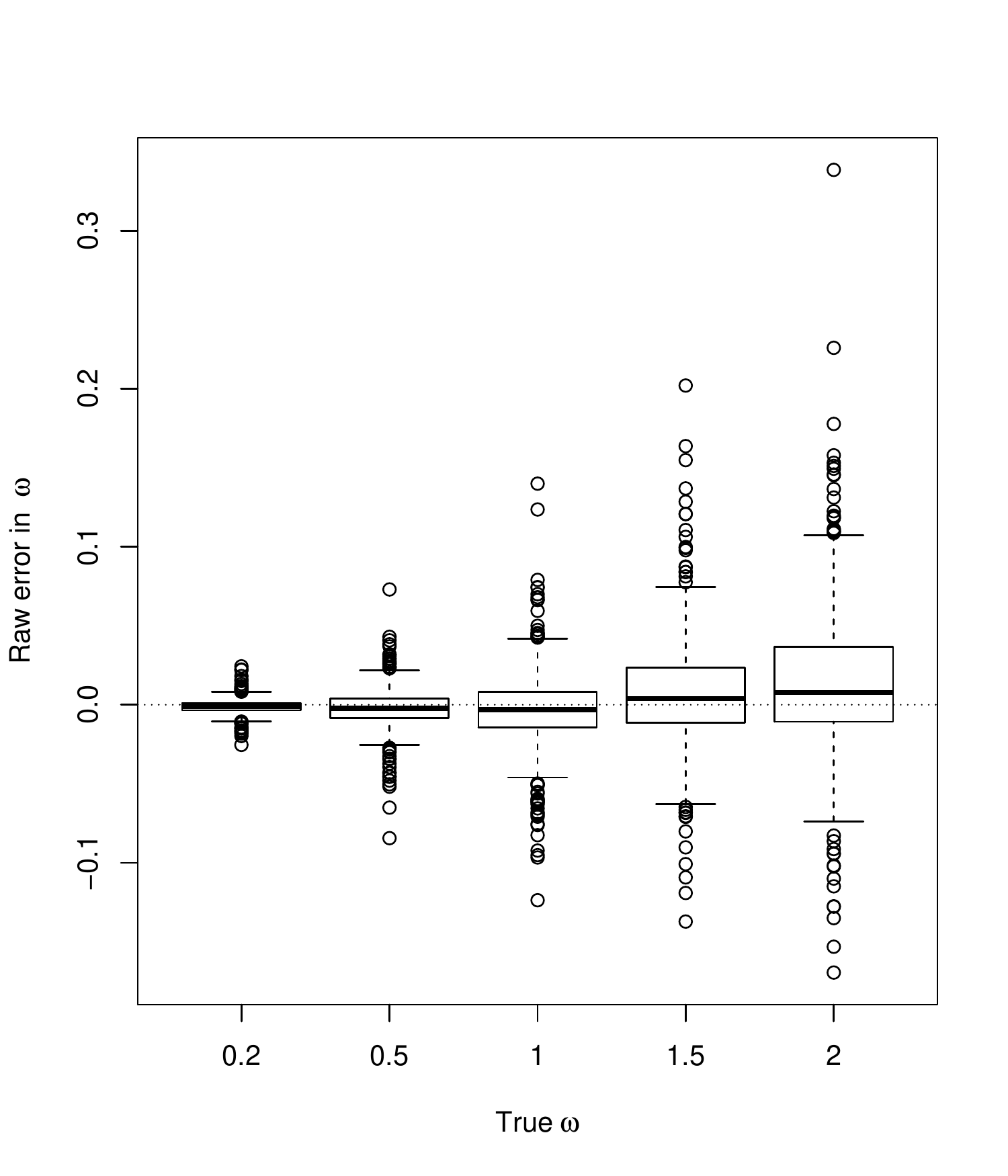}
& \includegraphics[width=0.49\textwidth]{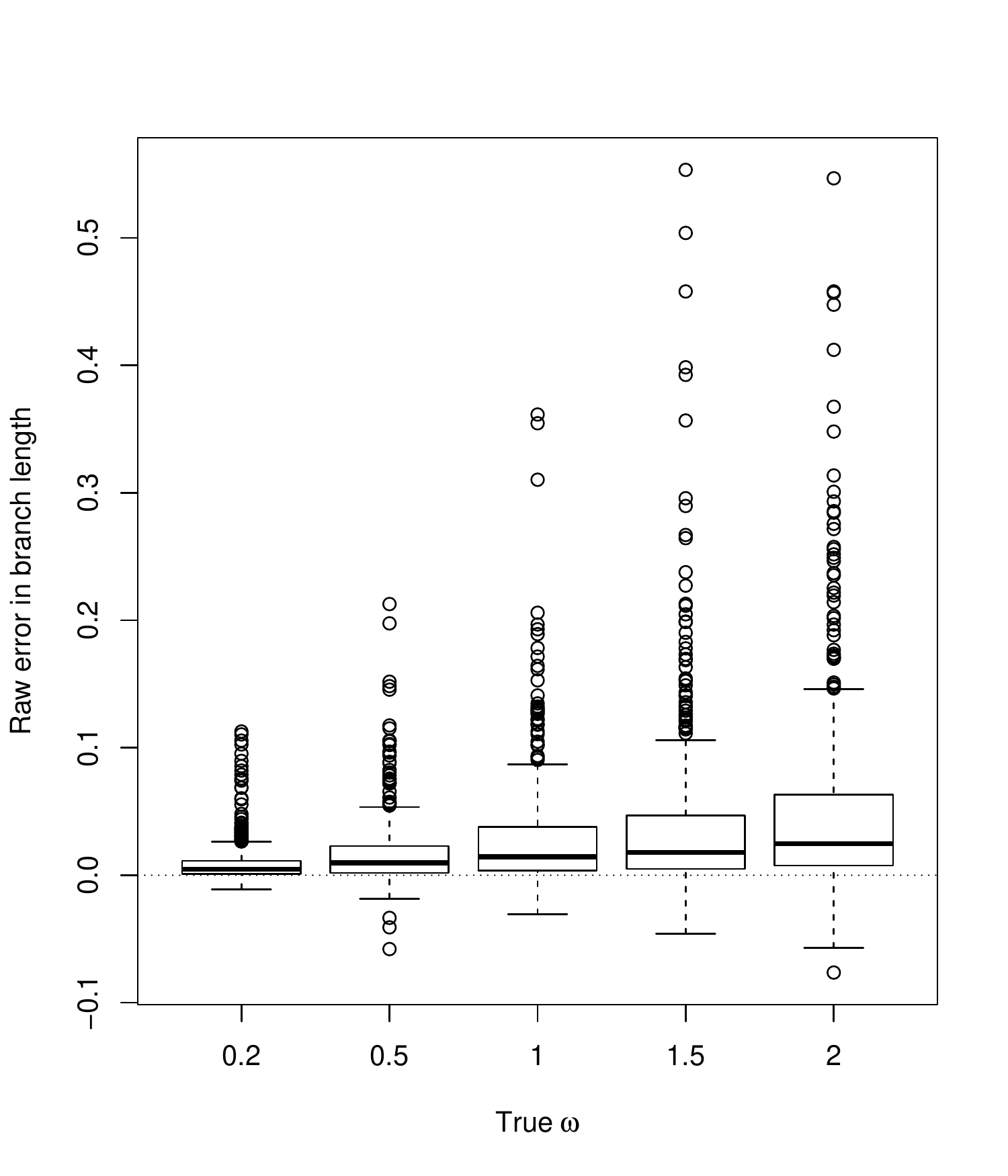} \\
(a) & (b)
\end{tabular}
\end{center}
\caption{Error in (a) $\omega$ and (b) branch length for the MG-style model that embeds the HKY DNA model. Boxplots display 500 simulations for each true value of $\omega$. Boxes show the lower quartile, median and upper quartile. Errors more than 1.5 times the interquartile range from the lower/upper quartiles are shown as individual points. 
Results are shown for simulations where the root distribution $\bm{\pi}_0$ was chosen at random.}
\label{fig:error_MGHKY_nasty}
\end{figure*}

Branch lengths are also mis-estimated (Figure~\ref{fig:error_MGHKY_nasty}(b), Table~\ref{tab:mean_error_bl_nasty}; Supplementary Figure~S2, Supplementary Table~S5). 
In the simulations where $\bm{\pi}_0$ was chosen to be intermediate, branch lengths are about equally likely to be over- or underestimated for values of true $\omega$ less than or equal to 1, and are more likely to be overestimated for true $\omega$ of 1.5 and 2. 
Mis-estimation is usually less than 1\% of true branch length but can be up to 8.5\%. 
For the set of simulations where the root distribution was chosen at random, branch lengths were far larger and also far more likely to be overestimated: for these experiments, the error was usually around 5\% but could be up to 83\% of the true branch length (Table~\ref{tab:mean_error_bl_nasty}).

\begin{table*}
\begin{centering}
\begin{tabular}{cccccc}
\hline 
true $\omega$ & 0.2 & 0.5 & 1 & 1.5 & 2 \tabularnewline
\hline 
Mean raw				& 0.010231 & 0.017138 & 0.028055 & 0.040812 & 0.050529    \tabularnewline
Mean relative error		& 0.047293 & 0.051239 & 0.052850 & 0.053294 & 0.052124  \tabularnewline
Maximum relative error	& 0.544113 & 0.514956 & 0.584356 & 0.730939 & 0.827849   \tabularnewline
\hline 
\end{tabular}
\par\end{centering}
\caption{Mean error, mean relative error and maximum relative error in the estimate of branch length for different true values of $\omega$ for simulations where $\bm{\pi}_0$ is chosen at random.
\label{tab:mean_error_bl_nasty}}
\end{table*}

As the underlying process becomes more heterogeneous the size of the errors increases. The largest errors in estimation of both $\omega$ and branch lengths occur when the difference in the rate matrices and/or difference in base frequencies are large (Figure \ref{fig:error_fanout_MGHKY_nasty}, Figure~S3).
	
\begin{figure*}
\includegraphics[width=\textwidth]{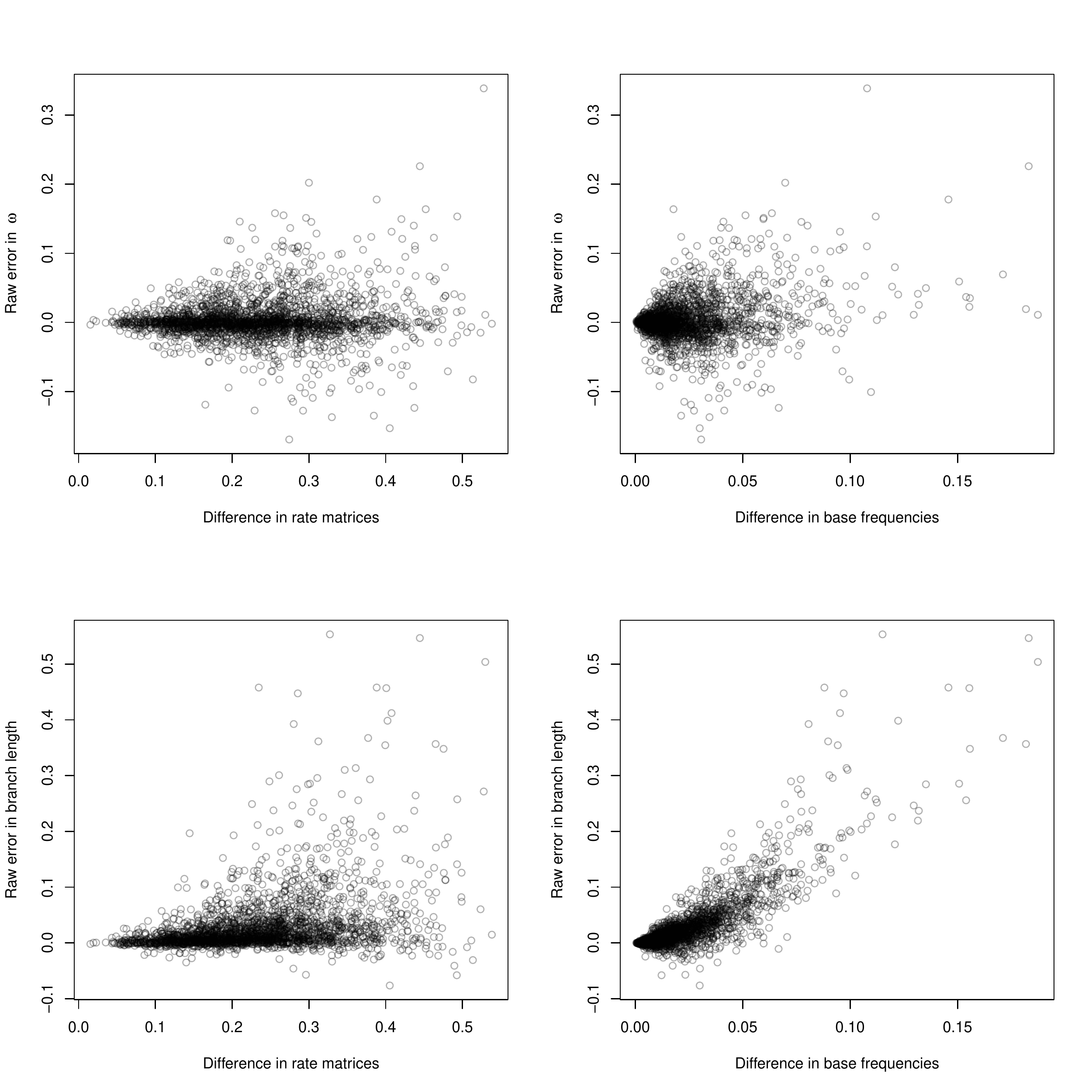}
\caption{Raw errors in $\omega$ (top panels) and errors in branch lengths (bottom panels)
for increasingly heterogeneous processes as measured by the difference in rate matrices (left-hand panels) and difference in base frequencies (right-hand panels).
Results are shown for simulations where the root distribution $\bm{\pi}_0$ was chosen at random. 
\label{fig:error_fanout_MGHKY_nasty}}
\end{figure*}

\subsection{Does choice of DNA model affect estimation of $\omega$ or branch lengths?}

As the results above indicate, lack of closure can cause mis-estimation of both $\omega$ and branch lengths. We were interested in whether the choice of underlying DNA model had an effect on accuracy. 
In particular we explored if using closed DNA models would reduce errors. 
To assess this we fit a linear model with relative error (in either $\omega$ or branch lengths) as the response variable and the following predictor variables: model class Lie-Markov \textit{vs.}\ non-LM),
number of parameters in the base model (treated as a scaled variable), number of base frequency degrees of freedom (treated as a categorical variable), difference in rate matrices, and difference in base frequencies. 
The relationship between the first three of these predictor variables and the models we used can be seen in Table~\ref{tab:model_families}.
We restricted to data sets where the true value of $\omega$ was 1.

\begin{table*}
\begin{centering}
\begin{tabular}{cllll}
\hline 
       Variable &           Estimate & Std.\ Error & t value & p-value  \tabularnewline  
\hline
(Intercept)       & -3.306e-03 & 1.646e-04 & -20.080 &  {\bf $\leq$ 2e-16} \tabularnewline
Model class (TR) & -1.333e-03 & 1.062e-04 & -12.551  & {\bf $\leq$ 2e-16} \tabularnewline
\# parameters         & 6.618e-04 & 2.472e-05 & 26.768 & {\bf $\leq$ 2e-16} \tabularnewline
BDF=1             & -2.972e-04 & 1.553e-04 & -1.913 &  0.0558     \tabularnewline
BDF=2            &-1.536e-04 & 1.678e-04 & -0.915  & 0.3602   \tabularnewline
BDF=3             &2.877e-05 & 1.908e-04 &  0.151  & 0.8802     \tabularnewline
diff.\ in rate matrices   &  1.691e-02 & 7.807e-04 & 21.661  & {\bf $\leq$ 2e-16} \tabularnewline
diff.\ in base frequencies   & 1.376e-01 & 6.917e-03 & 19.888 & {\bf $\leq$ 2e-16} \tabularnewline
\hline 
\end{tabular}
\par\end{centering}
\caption{Results of fitting linear model with relative error in $\omega$ as the response variable. 
With regard to the categorical variables, the intercept corresponds to models in the Lie-Markov (LM) class with BDF~=~0. As low errors are good, categories with negative coefficients in the `Estimate' column are performing well. In particular, TR (time reversible) models generate lower errors in $\omega$ than do Lie-Markov based models. 
\label{tab:lm_omega}}
\end{table*}

\begin{table*}
\begin{centering}
\begin{tabular}{cllll}
\hline 
       Variable &           Estimate & Std.\ Error & t value & p-value  \tabularnewline  
\hline
(Intercept)	& -2.508e-03  & 1.272e-04 & -19.714&  {\bf $\leq$ 2e-16} \tabularnewline
Model class (TR) & 8.547e-04 & 8.205e-05 & 10.418  & {\bf $\leq$ 2e-16} \tabularnewline
\# parameters         & 2.839e-04 & 1.910e-05 & 14.861 & {\bf $\leq$ 2e-16} \tabularnewline
BDF=1               & 1.611e-03 & 1.200e-04 & 13.423 &  {\bf $\leq$ 2e-16}     \tabularnewline
BDF=2            & -8.560e-04 & 1.297e-04 & -6.600 & {\bf 4.21e-11} \tabularnewline
BDF=3             &4.678e-04 & 1.475e-04  & 3.172 & {\bf 0.00151}    \tabularnewline
diff. in rate matrices   &   1.098e-02 & 6.033e-04 & 18.198  & {\bf $\leq$ 2e-16} \tabularnewline
diff. in base frequencies   & 2.697e-01 & 5.345e-03 & 50.452  & {\bf $\leq$ 2e-16} \tabularnewline
\hline 
\end{tabular}
\par\end{centering}
\caption{Results of fitting linear model with relative error in branch length as the response variable. 
With regard to the categorical variables, the intercept corresponds to models in the Lie-Markov (LM) class with BDF~=~0. The interpretation of this table is similar to the previous one, and we see that TR models are worse (positive estimate coefficient) than Lie-Markov based models.
\label{tab:lm_bl}}
\end{table*}

These results are in Tables \ref{tab:lm_omega} and \ref{tab:lm_bl}. In these tables, ``\# parameters'', ``BDF'', ``diff. in rate matrices'' and ``diff. in base frequencies'' are potentially confounding variables being accounted for.
The row of greatest interest is the ``Model class'' row, which shows how time reversible models perform relative to Lie-Markov based models.
For the $\omega$ relative error (able~\ref{tab:lm_omega}) the negative coefficient in ``Model class'' indicates that, other variables being equal, time reversible models give lower error than Lie-Markov based models.
For the branch length relative error, the converse is true --- Lie-Markov based models give lower errors than time reversible.
In both cases, the effect size is small (on the order of 0.1\%) so these differences are of little importance.
The results shown are for the case where $\bm{\pi}_0$ was chosen to be intermediate, however, the random $\bm{\pi}_0$ results are similar (time reversible models being better at estimating $\omega$ and worse at estimating branch length.)

\section{Discussion}

In this paper we introduced an algebraic formulation of codon models that allows us to separate the effect of independent evolution of sites across triplets (implemented via the Kronecker product) from the overlaying effect of the genetic code. 
This gives a flexible method for constructing codon models from any chosen DNA model (or even three different DNA models acting at different codon positions). 
By using this formulation, we can show that while closure properties of DNA models carry over to triplet models, the overlaying effect of the genetic code removes the closure property from codon models. 

Given that previous work suggests lack of closure could exacerbate misestimation of heterogeneous processes \cite{sumner2012general}, we used two-taxon simulations to investigate the effect of using homogeneous codon models to fit a heterogeneous process. 
Specifically, we investigated cases where the selection parameter $\omega$ was constant but where parameters of the underlying DNA model were different in different branches.  
We found errors in the estimates of both $\omega$ and branch length, these errors became larger on average as the processes on the two branches differed more. 
Further, where the ancestral codon frequencies were not intermediate, we encountered far larger errors than those where it was intermediate.
On average, the effect sizes are not large (less than 1\% for $\omega$ and less than 5\% for branch lengths), however, it was possible to get errors greater than 10\% for $\omega$ and greater than 50\% for branch lengths. 

\citet{kaehler2017standard} demonstrate that homogeneous time reversible (hence stationary) codon models are biased to overestimate $\omega$ when the sequences have differing codon frequencies. 
Our Monte Carlo simulations find this bias only when the true $\omega$ is strictly greater than 1.
We are simulating non-stationary non-homogeneous data and analysing it as stationary, homogeneous, but not necessarily time reversible. 
As our results when the underlying DNA model is non-reversible (Lie-Markov) are performing no better than time reversible DNA models, we can conclude that the misestimation of $\omega$ found by \citet{kaehler2017standard} is not due to the time reversibility assumption, but rather due to one or both of the stationarity or homogeneity assumptions.

Intriguingly, as pointed out in \citet{kaehler2017standard}, while non-stationary models pass absolute goodness-of-fit tests for nucleotide data, even the most general non-stationary model of codon evolution (GNC) is still frequently rejected by absolute goodness-of-fit tests. It seems plausible that the issues with lack of closure explored here may offer some explanation for the failure of these models to fit codon sequence data adequately.

To judge to what extent these errors were specifically due to lack of closure we repeated the simulations on triplet models with no $G$ matrix applied, in this case errors for both closed and non-closed DNA models are about an order of magnitude smaller (Supplementary Material). 
This result combined with the observation that codon models formed from closed DNA models (i.e. Lie Markov models) perform no better than models formed from other DNA models suggests that it is the lack of closure introduced by the genetic code that is the main source of the problem. 
This raises the question of whether there exist any biologically realistic codon models that are closed. 

This is currently an open problem. 
Mathematically, it appears to be a rather difficult task. 
The key issue is that applying augmentation parameters to allow the model to respect the genetic code introduces non-linearity into the model which, in itself, is enough to rule out the possibility of the model being closed.

A reasonable way around this difficulty is to ask, for a given non-closed codon model: ``What is the simplest closed codon model that contains this particular model as a submodel?''. The mathematical procedure for answering this question is computationally straightforward (albeit intensive), and our calculations have shown that the resulting models have many, many parameters (in the order of thousands) and are hence not of practical use for phylogenetics.
This does however prompt a modified question that has the potential to produce a more reasonable answer: ``What is the simplest \emph{linear} codon model that contains an MG-type model?'' In general a linear model is not closed, but our previous work \citep{sumner2012general} has shown that, at least in the case of DNA models, the errors caused by non-closure are, comparatively, resolved by moving from a non-closed non-linear model to a non-closed linear model.

For the MG-type models, the smallest containing linear model has quite an interesting structure, that we now describe.
To fix ideas, we discuss the MG model with F81 as the underlying DNA model and augmentation parameters 0, 1, and $\omega$.
To find the smallest linear codon model containing this model, one may proceed by finding the set of codon rate matrices obtained by taking sums of codon rate matrices from this model.
This results in the replacement of the substitution rates $(\alpha_i, \alpha_i\omega)$ --- which, as discussed above, exhibit non-linear constraints --- with an independent set $(\alpha_i, \mu_i)$.
This has the effect of removing the non-linear constraints on the model, at the expense of moving from a five parameter model to an eight parameter model.
This does however produce a linear codon model which is consistent with the genetic code structure and allows for the recoverability of \emph{multiple} analogues of the $dN/dS$ via the definitions:
\[
\omega_i\equiv \frac{\mu_i}{\alpha_i}.
\]
While there are intriguing possibilities inspired by signals from protein biophysics embedded in the genetic code through differing amino acid properties, we leave analysis and application of this model to future work.

\subsubsection*{Acknowledgement}

This research was supported by Australian Research Council (ARC) Discovery Grant DP150100088. 
The author's would like to thank Simon Whelan for suggesting the line of enquiry explored in this work after JGS's presentation at the Phylogenetics: New Data, New Phylogenetic Challenges Workshop, University of Cambridge 2011.

\bibliographystyle{natbib}
\bibliography{gy}
\end{document}